# Development of high-resolution timing technique in the past 25 years

J. Va'vra

SLAC National Accelerator Laboratory, CA 94025, USA

*Abstract* – This paper reviews recent progress in high-resolution timing, rate capability, and aging performance of particle detectors in the context of developments over the past 25 years. It focuses on MCP, SiPM, LGAD, 3D silicon, and SiEM detectors[1] and compares their performance with that of gaseous timing detectors. In addition, preliminary results from a simplified simulation program, Mini-TCAD, are presented, indicating that hexagonal 3D silicon cells may provide better drift-time uniformity and timing resolution than square cells.

## INTRODUCTION

This paper[2] is a follow up to a review paper from 2019 MPGD conference [1]. Both papers together will thus cover developments over the past 25 years.

Measurement of time has been on mind of many physicists for many years, of course. A typical detector was a combination of fast PMTs and scintillator tiles. Table 1 shows B. Attwood's review of status by 1980. At that time the best timing resolution was about 140 ps, achieved was by the Mark III experiment at SLAC. The point of showing this is that the next step in timing resolution in very large experiment was about ~80 ps in ALICE about 40 years later – see section 6. This shows that it takes a very long time to improve timing resolution significantly. I let the reader to predict when a similarly large experiment will operate routinely with a ~20-30 ps resolution.

**Table 1:** Status of TOF detectors by the year of 1980 [2]:

| Counter | L(cm) | $N_e$ | $\Delta\tau$(psec) |
|---|---|---|---|
| 1) MARK II[2] | 350 | 40 | 255 |
| 2) "Free Quark Search" (PEP-14) | 315 | 90 | 166 |
| 3) DASP[4] | 172 | 28* | 212 |
| 4) F. Binon et al., N.I.M., 153, 409 (1978) | 25 | 28* | 92 |
| 5) M. Wollstadt[7] | 100 | 39* | 144 |
| 6) M. Wollstadt[7] | 50 | 16* | 152 |
| 7) MARK III[5] | 300 | 120 | 140 |
| 8) M. Wollstadt[7] | 100 | 260* | 85 |
| 9) Same as 4) | ~.2 | 4500* | 48 |

25 years ago, a new effort to improve timing came because of introduction of new multi-pixel MCP and SiPM detectors, triggering new R&D motivated by of proposals for SuperB and Belle II experiments. Over past 25 years, there was a general push for higher luminosity not only at LHC but also at Belle-II, Electron-ion collider, Panda, etc. Table 2 shows examples of expected rates and fluences in the central detector regions for future CERN accelerators. We see that the future detectors must sustain enormous rates, and measure position to 20-30 μm and MIP time resolution to 20-40ps, and all this while power consumption must be limited to less than ~1.5 W/cm$^2$. It is obvious that not many detectors presented in this presentation will be able to run ~20 GHz/cm$^2$, and therefore a new detector R&D is needed, while realizing that the only way to cope with such rates is to reduce the detector pixel sizes, which, de facto, means Si-based detectors. This also means a great demand on the electronics development. That does not mean that other detector development is not important.

**Table 2:** Expected maximum rates and neutron equivalent fluences in future CERN accelerators in central tracking volume [3]:

| **Accelerator:** | **LHC** | **HL-LHC** | **SPS** | **FCC-ee** | **FCC-hh** |
|---|---|---|---|---|---|
| **Fluence [$n_{eq}$/cm$^2$/year]:** | few x $10^{15}$ | $10^{16}$ | < $10^{17}$ | $10^{10}$ | < $10^{17}$ |
| **Max hit rate [sec$^{-1}$ cm$^{-2}$]:** | 100 M | 2 - 4 G | 8 G | 20 M | 20 G |

This work supported by the Department of Energy, contract DEAC02-76SF00515.

[1] Detector abbreviations: MCP = microchannel plates, SiPM = Silicon PMT, LGAD = low gain avalanche diode, 3D Si = 3D Silicon, SiEM = Silicon electron multiplier.
[2] Invited talk at Micro-Pattern Gaseous Detector conference at Prague, Czech Republic, Aug 31, 2026.

# 1. Physics of Fast Timing Detectors

A timing technique seems simple on a piece of paper, but there are many hidden effects which have to be overcome. One can group them into these three categories:

a) **Signal formation (Shockley-Ramo theorem)** [4-5]:

Ramo has found that the detector current $i$ is proportional to electron charge $q$, weighting field $\overrightarrow{E_w}$, introduced by Shockley, and electron drift velocity $\vec{v}$. The weighting field is determined by the detector geometry and drift velocity $\vec{v}$ is determined by the electric field $\overrightarrow{E_{drift}} \neq \overrightarrow{E_w}$ :

$$i = q\,\overrightarrow{E_w}\,\vec{v} \tag{1}$$

We will see later in the discussion about Si 3D detectors that although the Ramo theorem is important, the final timing resolution is dominated by position-dependent drift times, stochastic ionization fluctuations, diffusion, and electronic noise. The relative importance of these contributions depends on the detector geometry and operating conditions.

b) **Detector timing resolution:**

One can start with a simple, somewhat naive formula in order to judge the timing resolution limit ($\sigma_{time}$):[3]

$$\sigma_{time} \sim \sigma_{noise}/(dV/dt)_{threshold} \sim t_{rise\text{-}time}/(S/N), \tag{2}$$

where V is the signal amplitude, N = $\sigma_{noise}$ is a noise. “One can say: Show me your pulses and noise level, and I will estimate your resolution.” For a typical MCP-PMT risetime of $t_{risetime}$ ~ 100 ps, one needs S/N ~10 to get into the ~10 ps regime. For a slower Si-detector risetime of $t_{risetime}$ ~2 ns, one needs S/N ~100 to get into the same timing regime. Gain in Si detector of 10 helps significantly.

c) **Final resolution:**

$$(\sigma_{Total})^2 \sim [\sigma_{Time}^2 + \sigma_{Landau}^2 + \sigma_{Diffusion}^2 + \sigma_{calibration}^2 + \sigma_{Clock}^2 + \sigma_{Chromatic\ effects}^2 + \sigma_{Laser}^2 + \ldots] \tag{3}$$

The $t_0$ time can easily dominate. There are many other obscure effects, which can affect timing resolution. For example: time walk, cross-talk effects in multi-pixel detectors, baseline ringing in multi-pixel detectors affecting later arriving pulses, charge sharing in multi-pixel detectors, chromatic effects, clock distribution throughout the system, pulse tail recovery, calibration, etc. The electronics resolution typically reaches ~2 ps [1], while the best overall is the DRS-4 electronics reaching <1 ps for short delay between start & stop [6]. For example, the chromatic error is important for DIRC-like detectors, which have a significant time of propagation difference between red and blue photons due to a long photon propagation in quartz [7].

# 2. Choice of Timing Method

In the past 25 years people used many methods [1]: (a) Leading-edge timing using TDC + ADC measurement to make pulse height correction, (b) timing with a hardware CFD circuits, such as LeCroy or Ortec 9327 CFD electronics (the hardware CFD circuit corrects for pulse height fluctuations and improves timing measurement considerably), (c) double-threshold timing, which was almost as good as CFD timing, (d) TOT timing, and (e) timing with waveform digitizers. Based on these experiences, my recommended timing procedure would be this:

a) **MCP single photoelectron** timing: use either a usual CFD method based on knowledge of digitized waveform, which normalizes pulses and applies 20% threshold to normalized pulses [1,8], or, simpler a double threshold timing.
b) **MCP MIP** timing: one can use similar procedure [9,10].
c) **Si-detectors MIP** timing: Double-threshold timing is an attractive option for LGAD and 3D silicon detectors. As discussed in Section 4.4, Mini-TCAD simulations indicate that it can outperform CFD timing for inclined tracks in 3D silicon cells. However, since detector can lose gain due to radiation, it is also important to keep track of it by measuring the total charge, which in turn can be used to adjust two thresholds as the detector gain changes.
d) However, generally speaking, the timing method depends on a choice of detector and should be studied in each case.

# 3. MCP Detectors

## 3.1 Ultimate MIP Timing Resolution with single pixel device

Microchannel plate detectors can achieve the best timing resolution presently. The ps-collaboration routinely uses Hamamatsu single pixel MCP for reference timing achieving ~5.1 ps timing resolution – see Fig.1. The detection scheme is based on a 20 GSa/s scope employing the CFD algorithm, and no amplifier [9,10].

One can ask if this can be improved further. If the detector would switch from Bialkali photocathode to the one with higher QE between 100-200nm (such as used for the Xe light detection), use $MgF_2$ window and use better digitizer providing less that 1 ps electronics contribution [6], one could, perhaps, reach 3-4ps for MIP timing.

[3] Well known formula in the communication engineering.

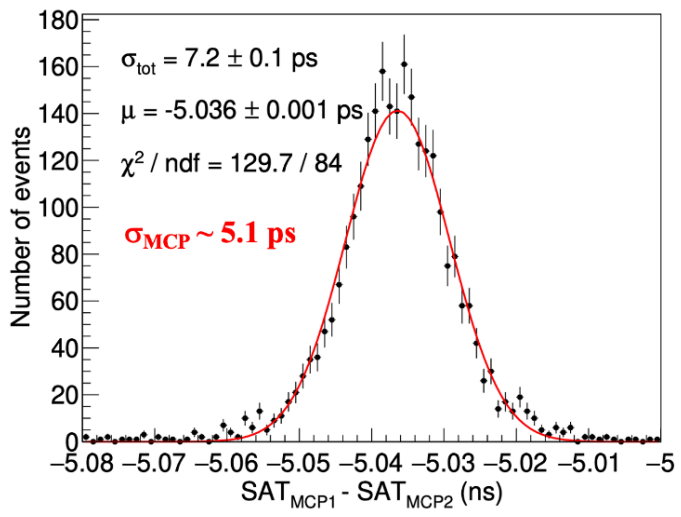

Operating point:

| | |
|---|---|
| MCP manufacturer | Hamamatsu |
| Fused silica window radiator | 3.2 mm |
| Photocathode | Bialkali |
| Number of pixels | Single |
| MCP Gain | ~ $8 \times 10^4$ |
| Total number of primary photoelectrons | ~ 44 |
| Total anode charge: | $3\text{-}4 \times 10^6$ el. |
| Detection method | 20GSa/s scope using CFD algorithm |
| Total charge per avalanche | $3\text{-}4 \times 10^6$ el. |
| Amplifier | No |
| Electronics resolution: 2.2 ps | ~ 2.2ps |
| Ion feedback | << 1% |
| Particles used for the test | Test beam |

**Fig. 1** The best single-pixel **MIP** timing resolution achieved so far using two Hamamatsu R3809U-50 MCPs [9,10] (left). The operating point during this test is shown on right.

## 3.2 Timing resolution in Multi-pixel HRPPD MCPs operating in Single Photon mode

US company Burle was to first company to bring multi-pixel MCPs, called Planacons, to HEP community at around 2000. Our group at SLAC was the first to use it on a large scale. SLAC group have done pioneering work studying these devices [1,10]. Burle MCPs were built by Jeff DeFazio, the same person who built Photonis MCPs when Photonis purchased Burle later. Photonis MCPs were used in Panda R&D effort and several examples will be given. However, Panda has terminated contract with Photonis due to production inconsistencies [11,12]. Electron Ion Collider (EIC) at Brookhaven Lab presently pushes their MCP effort around Incom's HRPPD[4] MCP to develop the DIRC and Aerogel RICH detectors for particle identification (PID). Fig.2 shows single photon timing resolution achieved by the team [13]. One can see that the core resolution is ~17 ps, which differs from RMS resolution, which is affected by electron recoils from top MCP. The rate of these recoils can be minimized by elevating top MCP-to-photocathode voltage [14], however, one has to be careful not to increase ion feedback, which, in my opinion, should be kept below 1%, even at the expense some detection efficiency loss. The choice of laser with very good timing resolution is very crucial to get excellent single pe timing resolution with a laser - see Fig.2 (right) and Ref.[39].

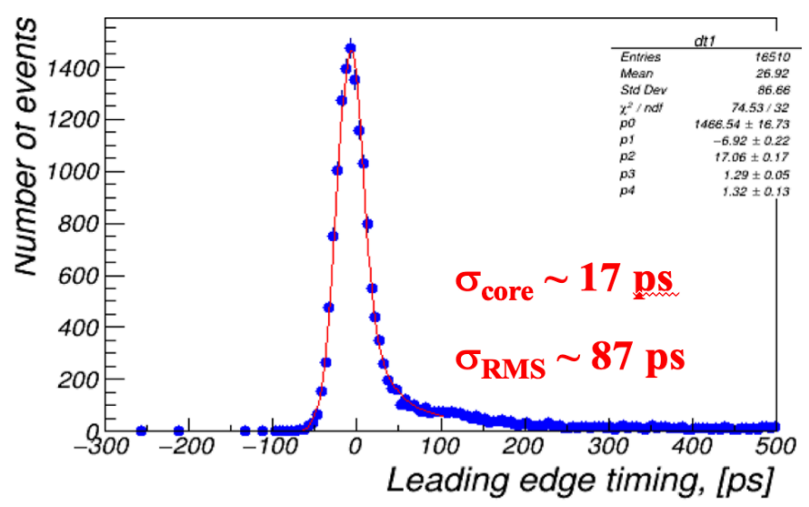

Operating point:

| | |
|---|---|
| MCP manufacturer | Incom |
| Total number of primary photoelectrons | 1 |
| MCP Gain | $\sim 1\text{-}2 \times 10^6$ |
| Photocathode | Bialkali |
| ALD coating | Yes |
| MCP pore dia. | 10 mm |
| Number of pixels | 1024 pixels, pixel size: 3 mm x 3 mm. |
| Amplifier | No |
| Detection method | Tektronix MSO66B scope (50 GS/s, 8 GHz analog bandwidth) with CFD algorithm |
| Laser type | Menlo Systems Model Elmo 780 (390 nm, σ <1ps resolution) |
| Coupling | DC coupled anodes |
| Dark noise | ~2 kHz/cm$^2$ at gain of ~$10^6$ |
| Cross-talk | <1-2%. |
| Ion feedback | ~2% |
| Particles used for the test | Laser |

**Fig. 2** The best multi-pixel **single-pe t**iming resolution achieved so far using Incom HRPPD (High-Rate Picosecond Photo-detector) [12]. Notice that the core resolution is different than the RMS resolution (left). Another thing to notice is a 2% ion feedback rate [13] at this operating point (right).

## 3.3 MCP maximum rate capability

Rate capability is becoming an important parameter when judging detectors. It is useful to recollect basic rate limits of the MCP. Fig.3a shows avalanche in one single MCP pore. Table on Fig.3c shows typical resistivity and capacitance on old MCPs, indicating that after each avalanche it takes 10-15 ms before the original potentials are re-established again, and the pore is ready for the next avalanche again with full gain. This limit is called the "**recharge limit**."

(a) (b)

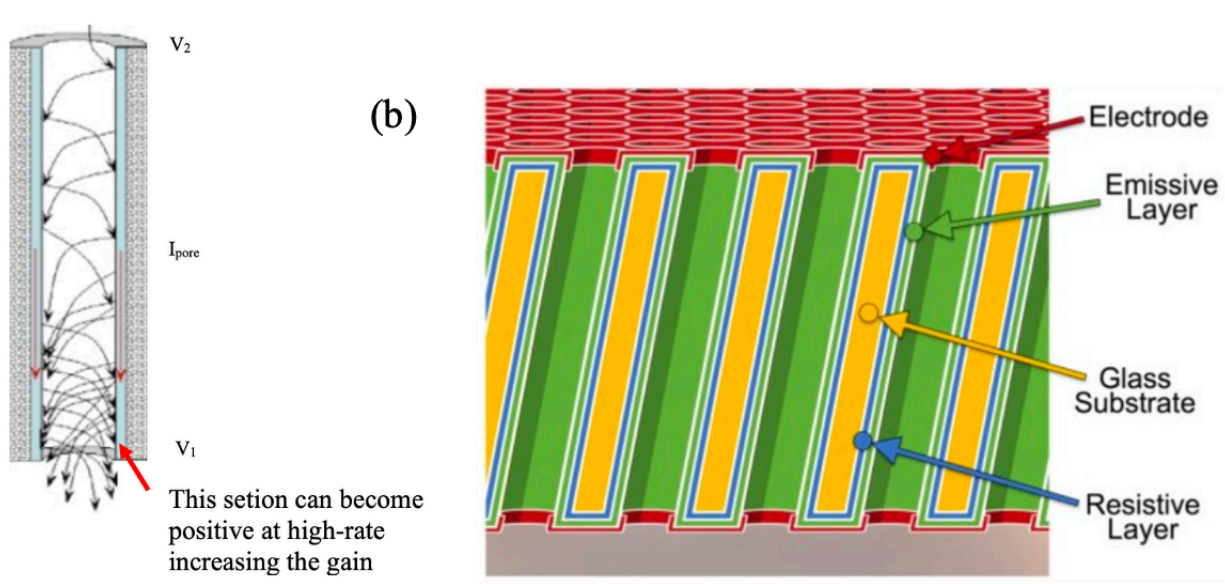

(c)

| Variable | Old MCPs | Low resistance MCPs |
|---|---|---|
| Typical pore resistance R: | ~few x $10^{16}$ Ω | ~2–3x lower |
| Typical pore capacitance: | ~$10^{-18}$ F | Slightly smaller |
| Recharge time constant RC: | ~ 10-15 ms | ~few ms scale |

**Fig. 3** (a) Avalanche in a single MCP pore; note that the end can end up slightly positive, which can increase gain at certain rate. (b) ALD coating of pores. (c) Typical recharge time constant of MC pore.

Recently, all companies started to treat MCP pores with Atomic Layer Deposition[5] (ALD) coatings – see Fig 3c. This treatment controls resistivity, which has some influence on the recharge limit, and the secondary emission, which controls the gain. However, table in Fig.3c shows that even ALD coated MCPs have a pore recharge limit in a ms range. Solution to the problem: (a) adding many small dia. pores in parallel reduces the probability of re-hitting the same pore resulting in detector's capability to handle higher rate, and (b) lowering resistance and lowering gain improves also the rate capability.

[4] HRPPD stands for High-Rate Picosecond Photodetectors MCP made by Incom Co.
[5] ALD coating was introduced by Irradiance Co. with collaboration of Argonne Natl. lab in USA. Not every company has the same recipe.

Fig.4 nicely illustrated the recharge limit using the Photonis MCP with 10 µm dia. pores, operating in single pe-mode with a gain of ~$10^6$ [12]. Fig.4b shows that focused electron beam, which can be enhanced by magnetic field, produces very low recharge limit. On the other hand, this effect can be reduced by diffuse illumination as shown on Fig.4c.

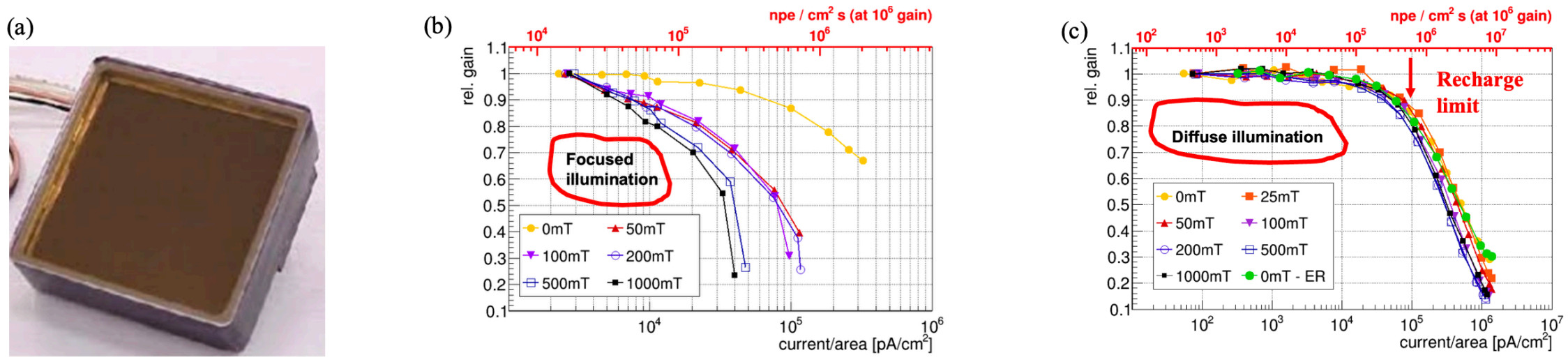

**Fig.4** (a) Panda Photonis MCP with ALD coating. [12] (b) Focused illumination → many electrons hit the same pores → early recharge limit (Magnetic field helps to focus electron beam). (c) Diffuse illumination → load shared across many pores → higher rate capability: ~400 kHz/cm$^2$ [14].

Fig.5 shows the effect of resistivity on the recharge limit [15], work done toward FP420 experiment. In this case, MCP operated at a gain of 2000 and used Photonis ALD coating. Fig. 5b shows that the best recharge limit depends on MCP resistivity, the best limit of ~20 MHz was achieved for resistivity of 27 MΩ, while still achieving excellent timing resolution – see Fig.5c. Notice a small gain increase for 27 & 35 MΩ, probably because of the effect mentioned on Fig.3a; for 44 MΩ, the recharge limit sets earlier than the gain increase. Laser tests indicate timing resolution of 20-30 ps up to rates of 30-40 MHz for 27 MΩ MCP resistance (Laser spot area not specified).

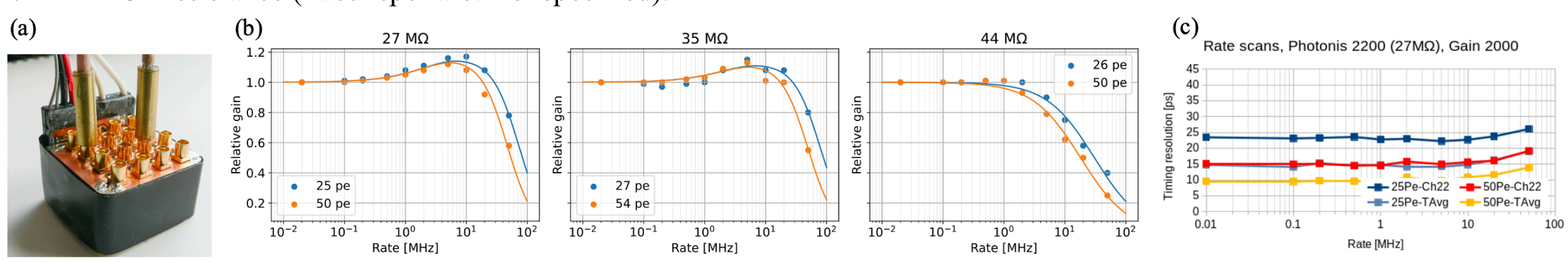

**Fig.5** (a) Photonis MCP used in FP420 experiment. Laser tests: (b) Gain dependence on rate for several MCP resistivities. (c) Timing resolution for the lowest resistivity MCP for several charges simulating MIP pulses [15].

On Fig.6, I summarize all MCP current density measurements, related to the recharge limit, on one plot. One can see that Komarek's and Melikyan's measurements at low gain, reach highest recharge limit in terms of 0.8-1.0 µA/cm$^2$ [16] If Melikyan would have chosen a similar gain as Komarek, it would limit higher recharge limit and he could have used ALD-coated MCPs.

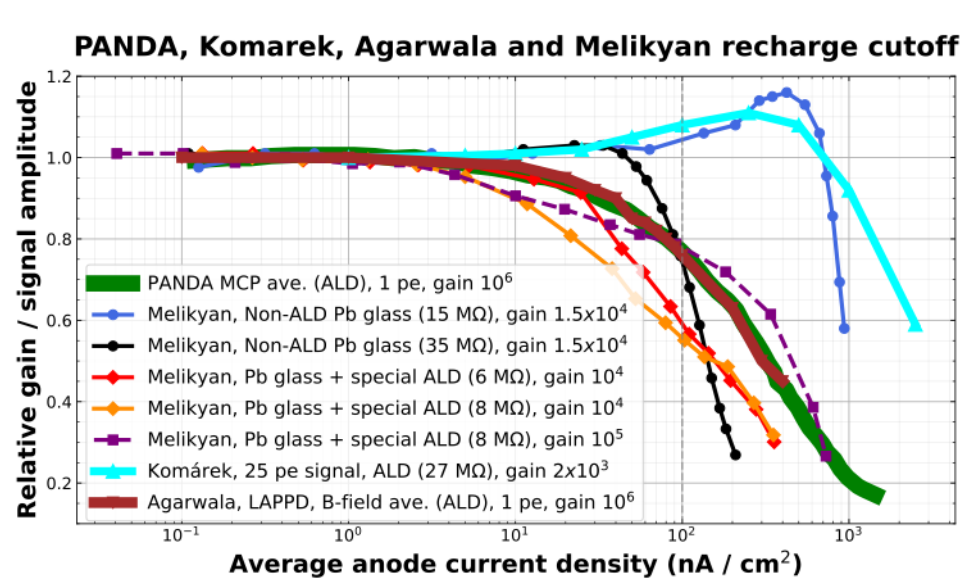

**Fig.6** The recharge limit is shown when plotting relative MCP gain as a function of average anode current density for several MCP tests operating either in single photon mode (gain ~$10^6$) or MIP detection mode with much lower gain.

### 3.4 MCP Photocathode Aging Tests

MCP detectors suffer from photocathode QE deterioration first before the gain is affected. The photocathode gets usually damaged by ion backflow. This means if we want to minimize this damage, we should reduce the gain to keep ion backflow at <1%. Fig.7 shows Panda tests of MCP photocathode aging by a photon flux. Photocathode lifetime reached accumulated charge of ~40 C/cm² @ MCP gain ~$10^6$. Since one cannot test every MCP for aging, they have tested only ~5% of all received. Making MCPs is still art to some degree; after a delivery of 60% of MCPs, Panda has terminated contract with Photonis due to inconsistencies in MCP performance [11,12].

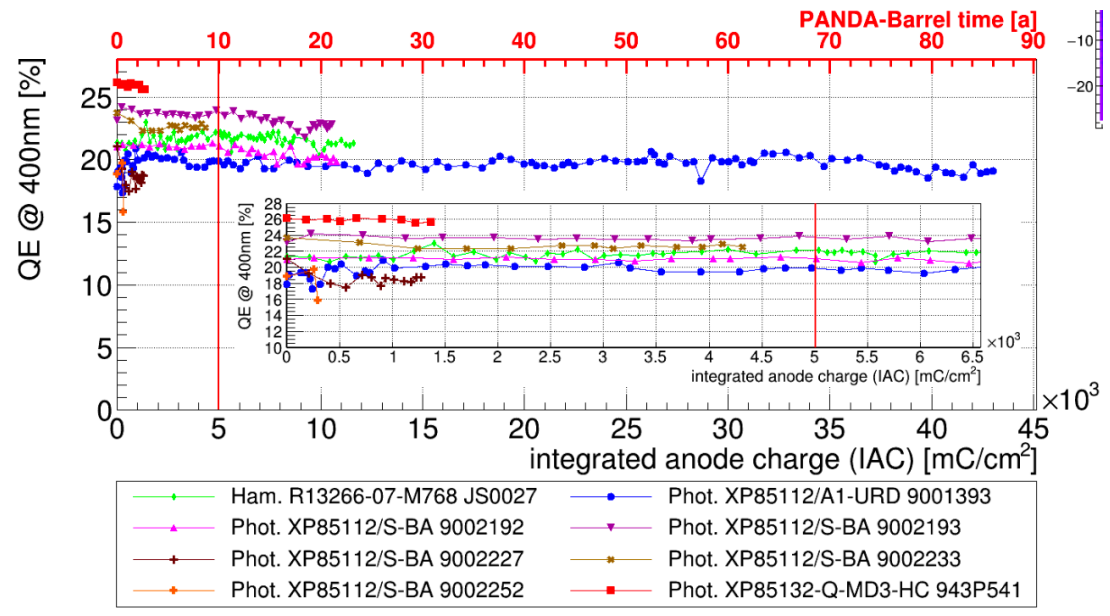

**Fig.7** Panda photocathode aging tests [Panda]. They have tested altogether 6 MCPs (~5% of total received), only 3 MCPs reached ~40 C/cm$^2$; 2 failed during aging tests (one due to a sudden escalation of noise, and one due to a sudden increase in the ion feedback production) [11].

### 3.4 Experience with running MCPs in real experiments

Belle-II experiment TOP DIRC detector is running since 2020. Fig.8a shows principle of this RICH detector, which used precise quartz bars as Cherenkov radiator producing photons, which propagate through internal reflection. The experiment measures time of photon propagation $T_{expected}$ -$T_{measured}$. Belle-II operates MCP at very low gain of ~2x10$^5$ to limit the ion backflow. In addition, they place Aluminum foil in front of the 2-nd MCP (see Fig. 8b) and they limit maximum background rate to ~850 kHz/cm$^2$ per MCP. Low MCP gain operation means to invest effort to higher electronics amplification, which in case of Belle-II running with gain of 110x.[6] Under these conditions, the Belle-II TOP counter is running with single photoelectron core resolution of $\sigma_{core}$ ~80 ps [17,18]. This is a different way than present way of running Panda prototypes, which runs gains at ~10$^6$ worrying about the loss of efficiency [11].

From Fig.8c we can that the experiment has been evolving as background conditions changed. As of writing this paper, they are still investigating a batch-to-batch dependency & reconciling bench aging tests and beam running [17].

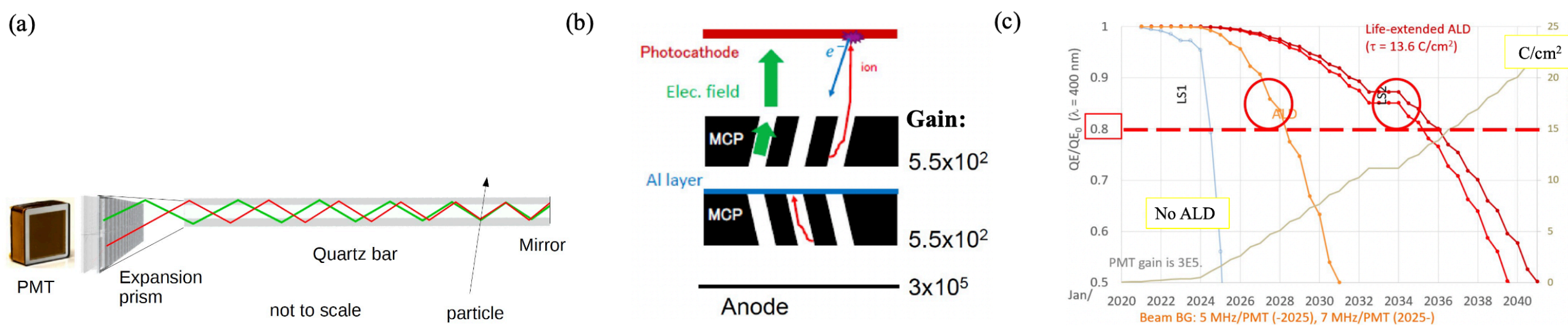

**Fig.8** (a) Belle-II DIRC with long quartz bars and HPK SL-10 16-pixel MCPs. (b) Present detector consists of two ALD-coated MCPs with similar gain/MCP, the 2-nd MCP has Al foil in front. The experiment is still tuning the gain in the 1-se MCP. (c) MCP gain drops when running high luminosity Belle-II (starting with no-ALD coated MCPs, to the 1-st generation ALD coating and ending with the present solution of ALD coating with Al-foil in front of the 2-nd MCP - Fig.8b) [17].

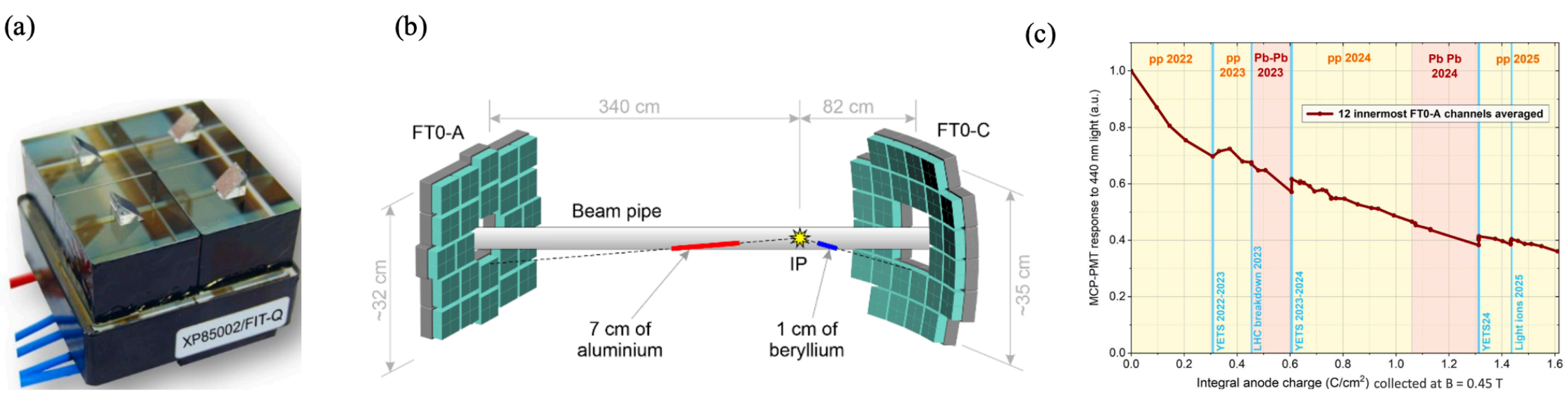

**Fig.9** (a) Planacon MCP with quartz radiator. (b) ALICE FIT detector. (c) Observed photocathode aging is substantial (~60% after ~1.6 C/cm$^2$) [16].

ALICE FIT experiment achieved excellent timing resolution ($\sigma$ ~17ps in p-p scattering, and $\sigma$ ~4.4ps in Pb-Pb collisions) [16]. This was achieved under a large background: MIP rate of 10-100 kHz/cm$^2$, ~3x10$^{10}$ $n_{eq}$/cm$^2$/month, ~4 krad of total ionizing dose per month, total program will have ~0.5 Mrad. To cope with such a background, the MCP gain was lowered to ~1.5x10$^4$, and MCP were selected to have a low resistance and no ALD coating. MCP recharge limit was close to 1 μA/cm$^2$. It would appear to me that a better choice would be to do what Komarek did, i.e., run even lower gain of 2000 and use ALD coating. Perhaps, under that condition the photocathode aging would be better than what is shown on Fig.9c.

To further improve MCP rate capability is to choose 3 MCP layers instead of two, which is a present choice of manufacturers. The 3-MCP choice suppresses the ion feedback, potentially improving QE stability at high rates. This was tested by Barnyakov et al. [19]. I believe that one could reach ~100 MHz/cm$^2$ by optimizing gain choices in various MCP layers.

[6] Belle-II is running MCPs similarly as the 1st FDIRC prototype (MCP gain ~2x10$^5$ with amplifier gain of ~130x) [1].

# 4. Si Detectors

## 4.1 CMS SiPM TOF Detector

Fig.10 shows CMS TOF detector based on LYSO bars and SiPMs [20]. The maximum rate capability is expected to be ~1.5 MHz/cm$^2$ and MIP timing resolution of ~25ps, which gets degraded to ~60ps after a dose of ~2x10$^{14}$ $n_{eq}$/cm$^2$. Their ASIC can measure dual threshold timing as well as total Q. To achieve a good timing resolution right from the beginning, they pay attention to calibration [21]. The issue with this detector is progressively increasing noise as SiPMs get radiation damaged, and it remains to be seen how the calibration will handle it. So, it is very important that this detector starts performing as TOF detector as soon as possible.

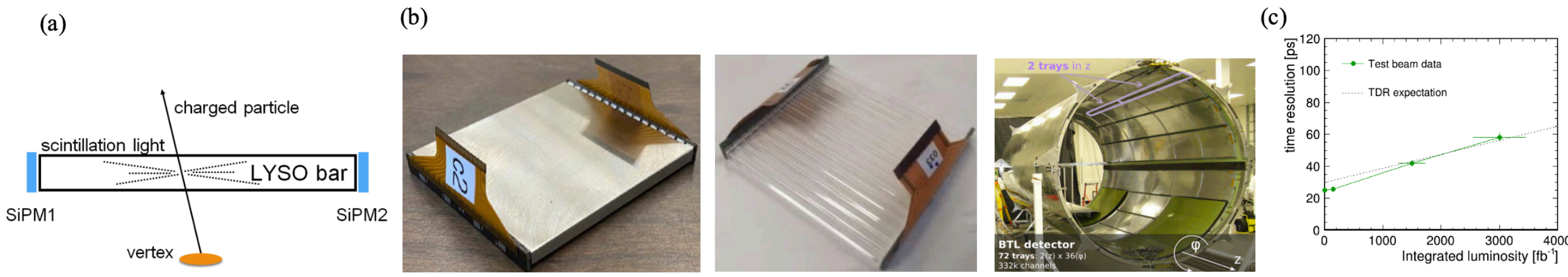

**Fig.10** (a) Principle of timing measurement in CMS TOF detector [20], employing LYSO crystals with two SiPMs at both ends. (b) Construction of a module out of 16 LYSO bars (54.7x3.12x3.75 mm$^3$, altogether 331,776 channels to calibrate. Operation assumes -35ºC, with occasional +60ºC annealing cycles. (c) Beam tests indicate MIP timing resolution of ~25 ps, which gets degraded to ~60ps after integrated luminosity of ~3000 fb$^{-1}$.

## 4.2 LGAD Detector

Fig.11 shows Low-Gain-Avalanche_diodes (LGAD), which is a Si detector with a gain of 10-20, which allows to reach S/N ratio of 15-30. The detector will operate at -30ºC using $CO_2$ cooling. In ATLAS application, the detector pixel size is 1.3x1.3 mm$^2$ and thickness of 50 µm [22, 23, 24]. The measured timing resolution of 35-40 ps was achieved, which is stable up to ~3x10$^{15}$ $n_{eq}$/cm$^2$; this resolution can be improved to ~30ps by thinning the sensor from 50 µm to 30 µm thickness; this works because late arriving charge does not contribute to timing. Fig.11c indicates the problem with this type of detector: the doping layer, responsible for the gain, gets damaged by radiation (the neutron limit is ~2.5x10$^{15}$ $n_{eq}$/cm$^2$ and proton limit is ~10$^{15}$ protons/cm$^2$ – after these limits are reached, the detector has to be replaced). With sufficiently small pixels (10–20 µm), rates of 20–40 GHz/cm² may be achievable in principle, provided that charge sharing, cross-talk, sensor recovery, and readout limitations can be controlled. However, such rates have not yet been experimentally demonstrated.

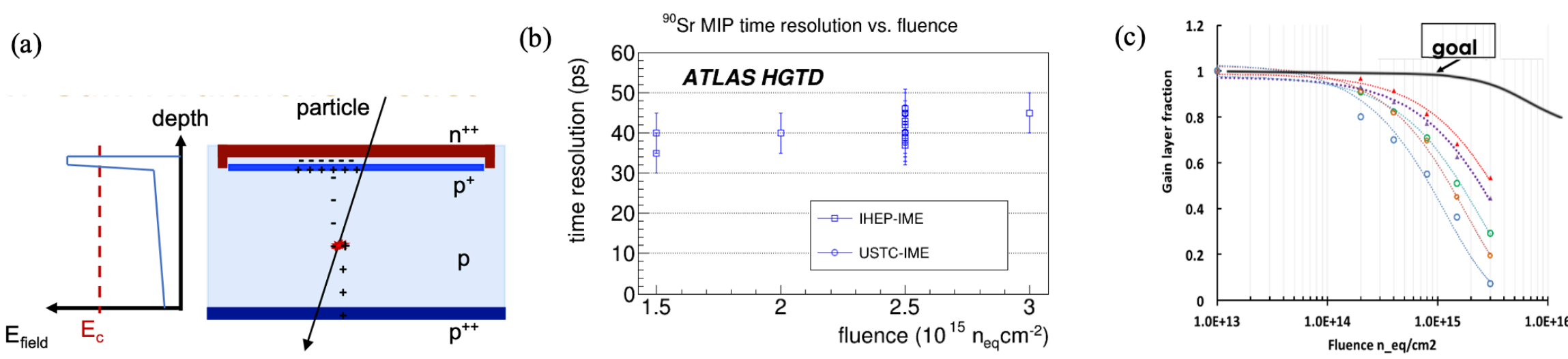

**Fig.11** (a) Principle of LGAD Si detector with a gain of 10-20. (b) Timing resolution as a function of fluence. (c) LGAD gain as a function of fluence.

## 4.3 Silicon Electron Multiplier (SiEM) Detector

Fig.12 shows a new concept of SiEM detector [25,26], which can provide gain (4x initially, but hopefully it will improve) without using doping mechanism as used in LGADs. Embedding metal electrodes inside the Si bulk create a tunable high-field region **without doping** as LGADs have. Fig.12a shows one of several possible initial designs. Expected maximum rate capability will be similar to LGAD (20-40 GHz/cm$^2$) assuming 10-20 µm pixelization. Expected timing resolution is $\sigma_{LGAD}^2 + \sigma_{pitch\ effect}^2$. The pitch contribution is shown on Fig.12c. Again, the detector should be thin as in the case of LGAD. Presently R&D is in progress, but no actual test beam data are available.

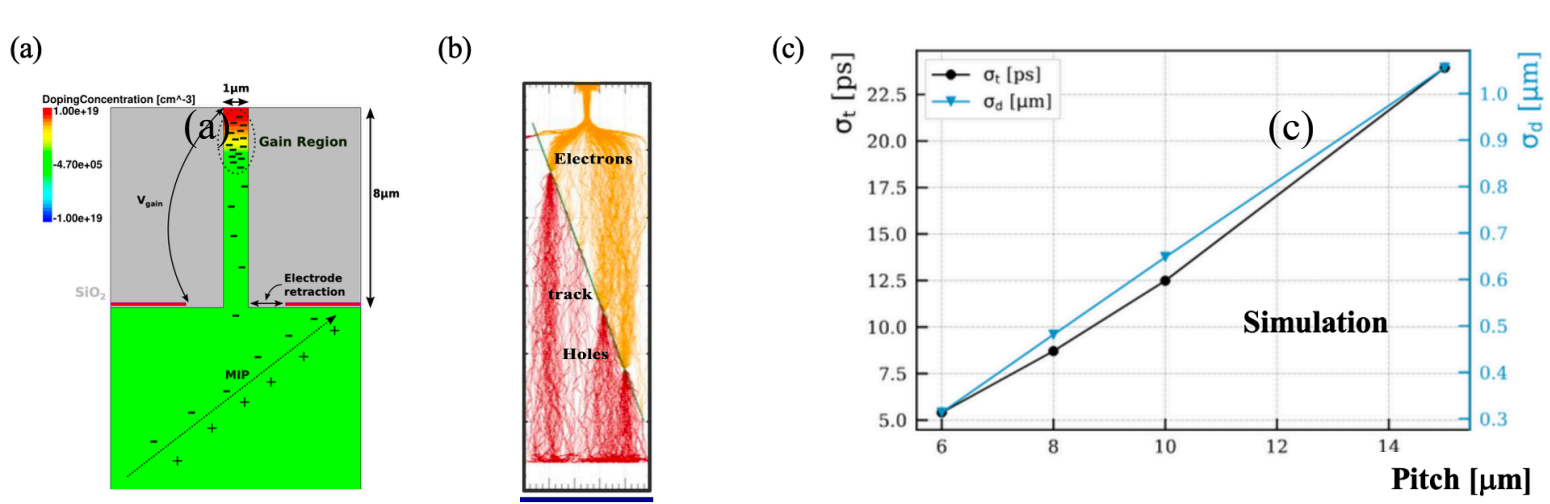

**Fig.12** (a) The most recent electrode arrangement of SiEM detector [25,26]. (b) Simulation of electron and hole drift in SiEM detector. (c) The expected timing resolution in SiEM is expected to be similar to LGAD detector plus a contribution from pitch effect shown on this figure.

### 4.4 Si 3D Detector

Fig.13a shows original concept of 3D Si detector proposed by S. Parker about 30 years ago [27]. It was resembling Si drift chamber. The TimeSpot collaboration came up with a concept of so called "trench" concept shown on Fig.13b [28]. The detector had a gain of 1. The prototype was subject to radiation damage of $10^{17}$ 1-MeV $n_{eq}/cm^2$, and still could produce timing resolution of ~11.5ps in test beam – see Fig.14a. However, as can be clearly seen on Fig.14b, the S/N ratio in this detector is rather poor.

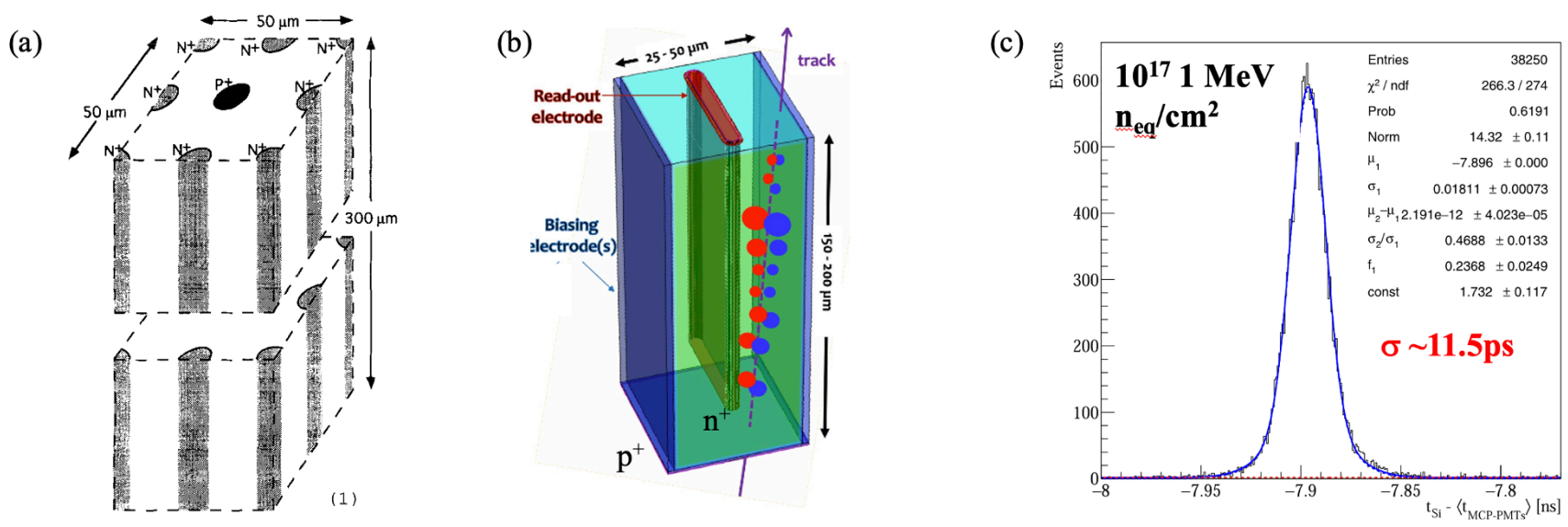

**Fig.13** (a) S. Parker's original concept of 3D detector [27]. (b) The TimeSpot collaboration's 3D "trench" cell [28]. Measured timing resolution with the TimeSpot "trench" cell detector, which was previously irradiated with ~$10^{17}$ 1 MeV $n_{eq}/cm^2$[28].

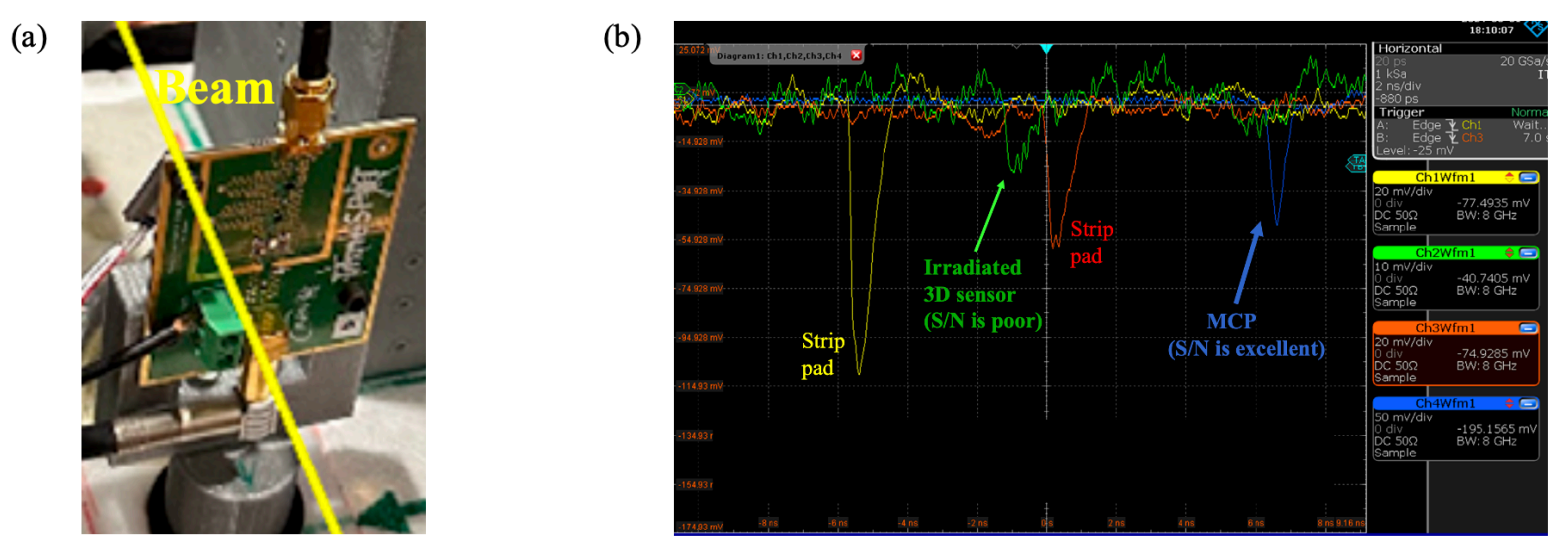

**Fig.14** (a) Test beam arrangement to test the "trench" 3D Si detector by TimeSpot collaboration [28]. (b) The scope picture shows that the previously irradiated 3D Si "trench" detector has poor S/N ratio compared to, for example, MCP detector used to provide a start signal [28].

To improve the S/N ratio in 3D Si detector, there is an effort to thin anode electrode to ~0.5 μm diameter. Fig.15a shows a possible concept, resembling a tube wire chamber in old days, proposed by Kramberger [29], who thinks that a moderate gain of 5-7 is possible. Presently, intensive R&D is in progress.

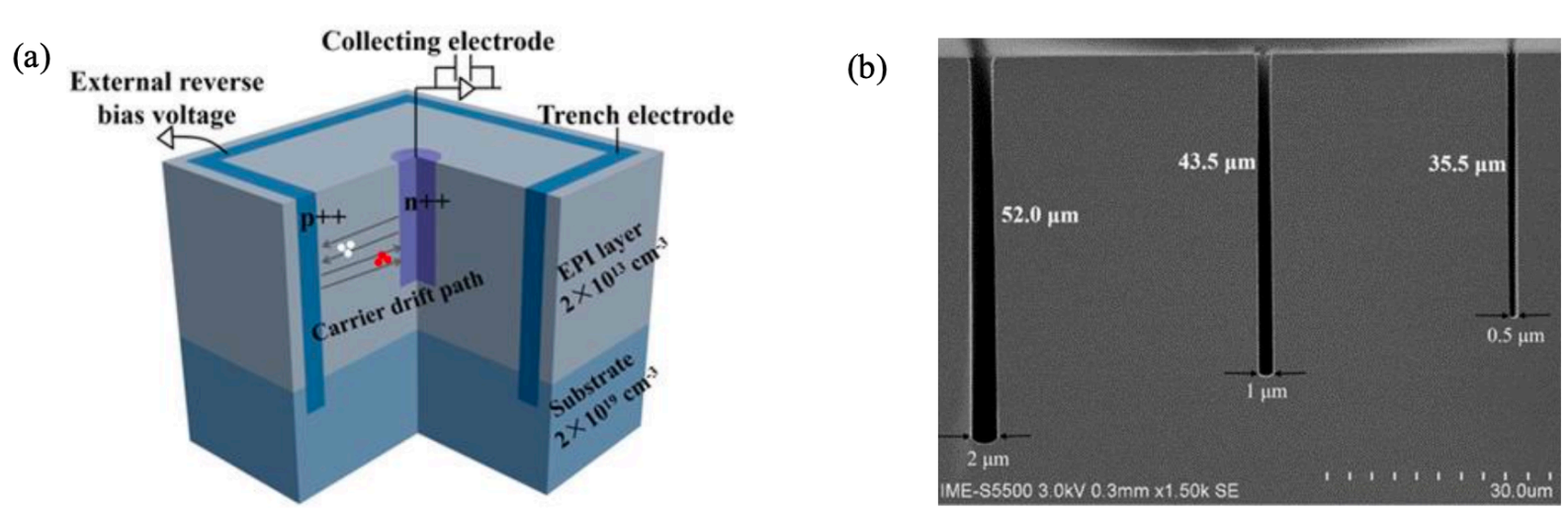

**Fig.15** (a) Square geometry concept of 3D Si detector as proposed by Kramberger [29]. (b) This indicates that a very small diameter close to 0.5 μm can be etched in Si [30].

I have created a drift program ~45 years ago [31] and decided to do a similar simulation for 3D Si detector [32], now with help of ChatGPT. In particular I wanted to know if a hexagonal geometry would be better choice. The Python code, called "Mini-TCAD", included: (a) calculate the two-dimensional electric field and electron drift trajectories, assuming no dependence on the third coordinate; (b) construct detector signals from the simulated charge arrival times, including amplifier response, (c) include ionization Landau-like fluctuations, (d) assume 0.5 μm anode providing gain of 7x, (e) assume hexagonal or square cells in x-y plane with 15-40 μm pitch and 50 μm z-depth, and (f) include amplifier shaping and electronics noise. The aim was to investigate whether hexagonal cells provide better timing resolution than square cells. Figs.16a&b show drift time map in the two types of cells and Fig.16c define the pitch. It is clear that square cells have regions near corners with very slow drift. Results of this simulation are shown on Fig.16d for tracks parallel to anode electrode ($0^o$). We find that for this case, CFD and double-threshold timing give similar timing resolution and hexagonal cells consistently give better timing resolution than square cells for all pitch distance studies. Both methods show increase in the timing resolution as a function of pitch. However, for inclined tracks (25-$30^o$), the double-threshold timing is better than CFD timing (RMS 29.7 ps (CFD) vs. 21.8 ps (double threshold) with 98% efficiency). The better performance of double-threshold timing for inclined tracks may be attributed to variations in pulse shape caused by late-arriving charge, particularly from regions near the cell corners. These variations can affect the CFD timing determination. Simulations indicate that position-dependent drift times, ionization fluctuations, and diffusion contribute to the timing resolution.

(a) (b) (c) (d)

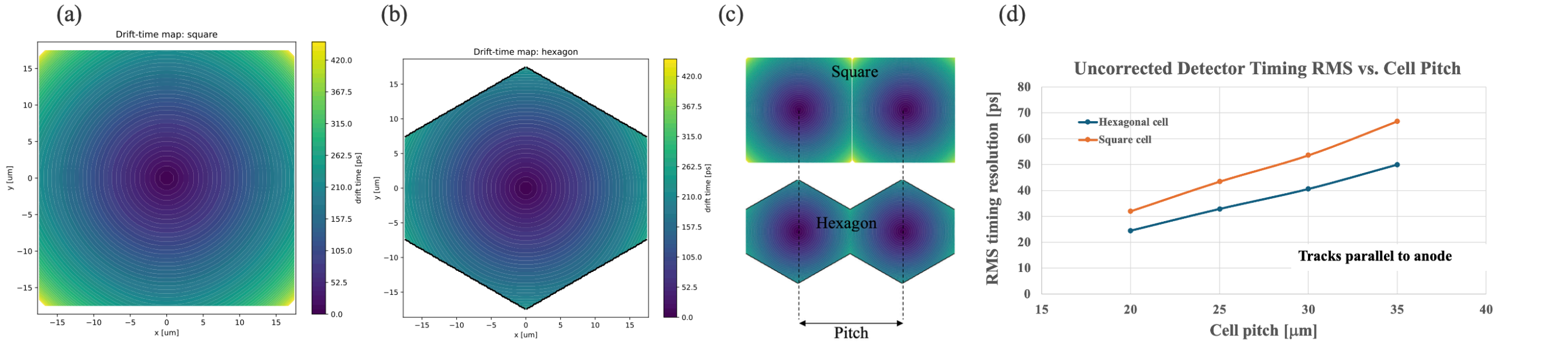

**Fig.16** Drift time map in (a) square cell, (b) hexagonal cell. (c) definition of pitch in two types of cells (sizes in microns). (d) Timing resolution as a function of pitch for both types of cells (in this case, tracks are parallel to anode electrode) [32].

# 5. Gaseous Detectors

## 5.1 Micromegas

Fig.17a shows excellent timing resolution result with Micromegas detector [34], operating with $Ne/CF_4/C_2H_6$ (80/10/10) at 1 bar. These are important points to mention: a) large extraction field shortens the distance to the first avalanche and reduces longitudinal diffusion, b) there are no focusing effects, c) UV photocathode & $MgF_2$ window produce larger photoelectron yield, d) the second gap mainly boosts the gain. It is essential to extract photoelectrons with very high field of 30-40 kV/cm, which lowers longitudinal diffusion. Drift cell is separated by mesh, held at ground potential, separating two drift field regions; one closer to pixels has a lower field. The total gain is in range of $10^5$-$10^6$, perhaps, too high from point of view of QE aging.

Excellent timing resolution of σ ~11 ps was achieved with CsI photocathode, which is an excellent match to VUV Cherenkov photons provided by $MgF_2$ radiator. One issue with this geometry and CsI photocathode is that there is no protection from the large ion feedback, which is enhanced by a large extraction field near photocathode. We can see on Fig.17d that a QE declines after a charge dose of only a few mC/cm$^2$, based on aging study using the same gas and gain [34]. Unfortunately, as the extraction field increases the timing resolution improves but the ion bombardment of photocathode increases. That is why there is an ongoing search for possible new photocathodes. So far only σ ~25-30 ps resolution was achieved DLC (Diamond-Like Carbon), Ti, B or C photocathodes, because they do not provide as many photoelectrons as CsI; search for new photocathodes continues. The aging limit with DLC photocathode is close to 100-200 mC/cm$^2$, if we accept a 10-15% QE decrease. These aging measurements were performed separately from the best timing measurements, and the photocathode lifetime under identical operating conditions remains to be established.

(a) (b)

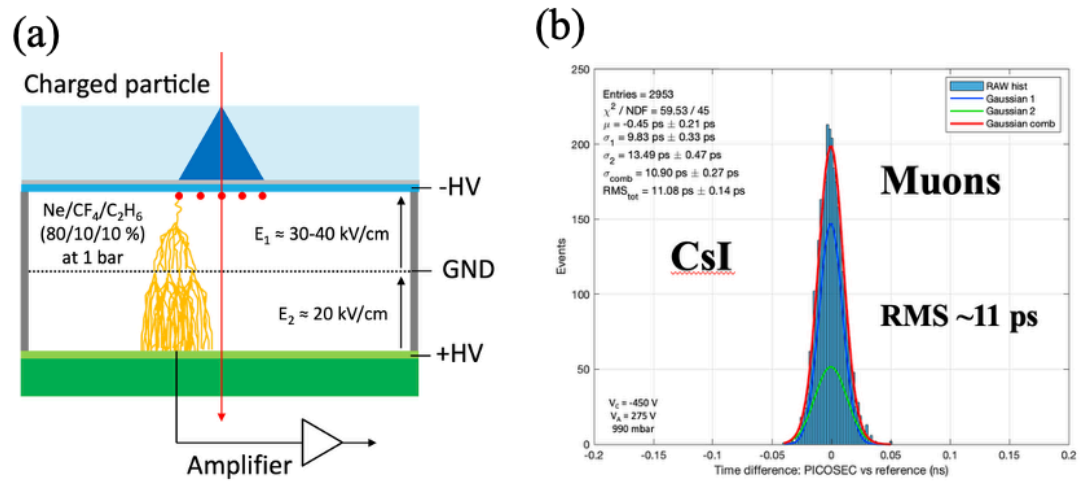

(c)

| | |
|---|---|
| Total gain | $10^5$-$10^6$ |
| Gas | $Ne/CF_4/C_2H_6$ (80/10/10) |
| Photocathode | Cr (3nm) + CsI (18 nm) |
| Preamplification gap | 120-200 μm |
| Amplification gap | 127 μm |
| Radiator | 3 mm thick $MgF_2$ |
| Pre-amplification field | 30-40 kV/cm |
| Amplification field | 20 kV/cm |

(d)

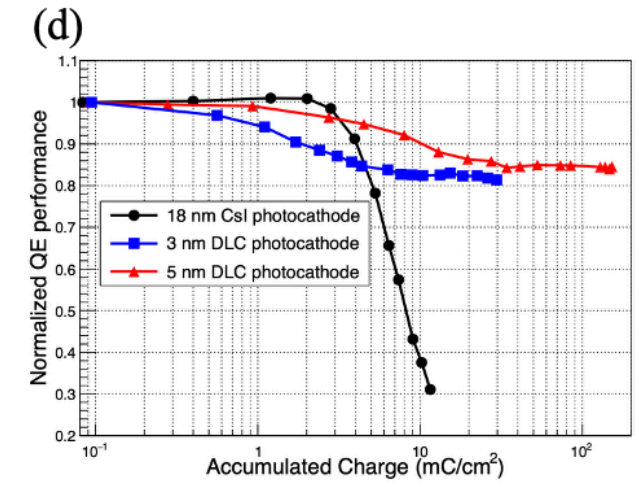

**Fig.17** (a) Principle Micromegas used for timing measurement [33]. (b) Timing resolution obtained with CsI photocathode and muon beam. (c) Table shows the operating point. (d) Aging test with the same gas but different photocathodes [34].

## 5.2 μ−RWELL

Figs.18a&b show details of μ-RWELL detector [35]. We notice that some portion of photo-electrons need to make a focusing turn to get to anode. This focusing effect will worsen the timing resolution compared to Micromegas, where electron paths are more direct. Indeed, Fig. 18c shows that this detector gets timing resolution of σ ~ 23.5 ps with CsI photocathode, a factor of ~2 worse than what was obtained with Micromegas [33] – see Fig. 17b. This resolution worsens to σ ~35 ps with a DLC photocathode due to lower number of photoelectrons. In a separate test, a different-geometry of μ-RWELL detector operating with DLC photocathode has reached MIP rate of ~10 MHz/cm$^2$ [36]; this test, however, measured neither timing resolution nor aging. The photocathode aging performance may be similar to that of Micromegas, assuming comparable ion backflow, but this remains to be demonstrated experimentally.

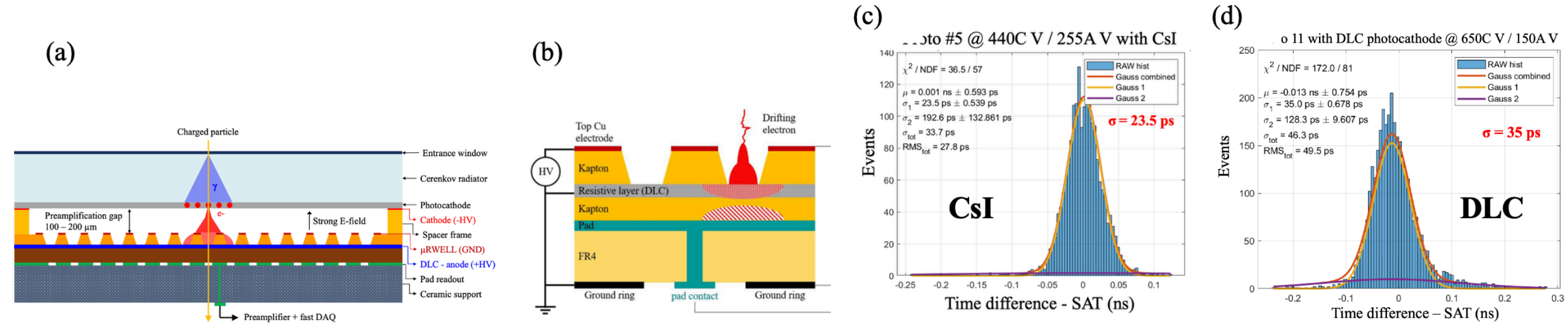

**Fig.18** (a & b) Principle of μ-RWELL detector [35]. (c) Timing resolution with CsI photocathode. (d) Timing resolution with DLC photocathode.

## 6. ALICE TOF Detector

It is good to end the paper with ALICE MRPC TOF detector – see Fig.19a. Fig.19b and 19c show the excellent particle identification at 13.6 TeV [38]. This experiment proved that one needs to pay attention to calibration issues before the experiment starts.

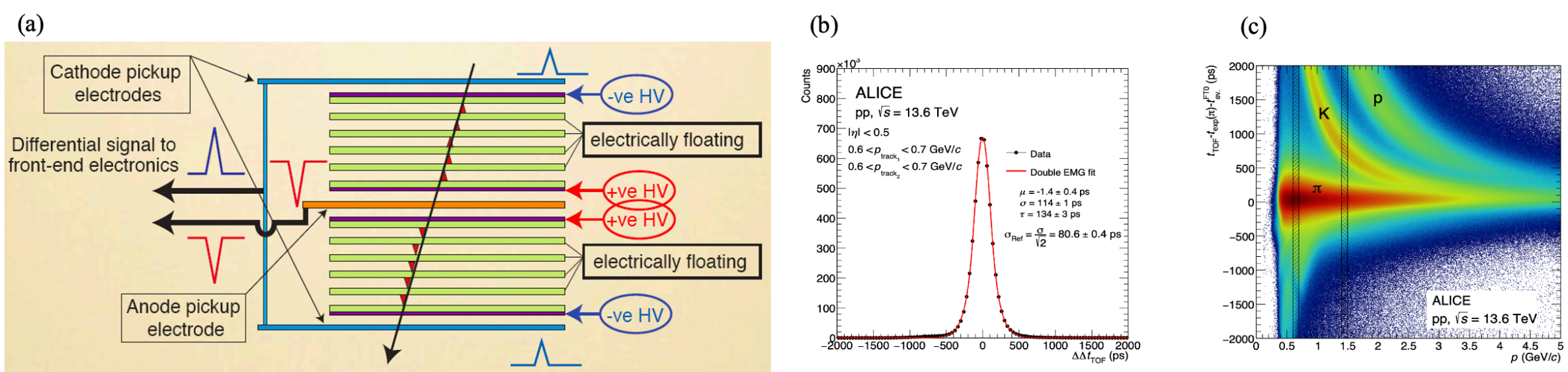

**Fig.19** (a) ALICE MRPC detector. (b)Timing resolution of ~80 ps achieved at ALICE. (c) Excellent particle identification by ALICE TOF system [38].

## 7. Summary of timing resolution, rate capability and aging tests

**Table 3:** Timing resolution, maximum rate capability and anode charge doses, obtained in beam tests or laser tests, (some numbers are either not known to author presently or not yet measured).

| Detector type | Experiment or beam test | Maximum rate | Maximum charge dose | Timing resolution (measured) | Ref. |
|---|---|---|---|---|---|
| MRPC | ALICE | 500 Hz/cm$^2$ *** | - | 80 ps *** (MIP) | [38] |
| Diamond | TOTEM | ~3 MHz/cm$^2$ * | - | ~ 80 ps/track/single sensor * (MIP) | [37] |
| MCP (single pixel) | ps collaboration | - | - | **5.1 ps** * (MIP) | [9,10] |
| Incom HRPPD MCP | BNL EIC | - | - | $\sigma_{core}$ ~**17 ps** & $\sigma_{RMS}$ ~87 ps * (single pe) | [13] |
| Photonis MCP | PANDA | 400 kHz/cm$^2$ * | **<40 C/cm$^2$** * | $\sigma_{core}$ ~26 ps & $\sigma_{RMS}$ ~109 ps * (single pe) | [11,12] |
| Hamamatsu MCP | Belle-II | **850 kHz/cm$^2$** *** | **13 C/cm$^2$** *** | $\sigma_{core}$ ~80 ps *** (single pe) | [17] |
| Photonis MCP | ALICE FIT | **0.8-1 μA/cm$^2$** *** | 1.6 C/cm$^2$ (65% QE drop) | 17 ps (p-p), **4.4 ps (Pb-Pb)** *** (MIP) | [16] |
| Photonis MCP | ATLAS Forward TOF | **0.8-1 μA/cm$^2$** * | - | 20-30 ps * (MIP) | [15] |
| 3-MCP | Barnyakov et al. | **100 MHz/1.8cm$^2$** * | - | - | [19] |
| LGAD (30-50 μm) | ATLAS | 20-40 GHz/cm$^2$ **** | 2.5x10$^{15}$ n$_{eq}$/cm$^2$ * | 30-50 ps * (MIP) | [22-24] |
| SiEM (Si-GEM) | ps collaboration | 20-40 GHz/cm$^2$ **** | 10$^{17}$ n$_{eq}$/cm$^2$ **** | - | [25,26] |
| 3D trench Si-sensor (no gain) | TimeSpot collaboration | <12 GHz/cm$^2$ ** (Limited by ASIC) | 10$^{17}$ n$_{eq}$/cm$^2$ * | **11.5 ps** (MIP) * | [28] |
| SiPM with LYSO | CMS TOF | 1.5 MHz/cm$^2$ * | 2 x 10$^{14}$ n$_{eq}$/cm$^2$ ** | 25 → 60 ps * (MIP) | [20] |
| Micromegas with CsI | ps collaboration | - | 2-3 mC/cm$^2$ ***** | **11 ps** * (MIP) | [33] |
| Micromegas with DLC | ps collaboration | 1-2 MHz/cm$^2$ ***** | 200 mC/cm$^2$ ***** | 35 ps * (MIP) | [34] |
| μ-RWELL with CsI | ps collaboration | - | - | 25-30 ps * (MIP) | [35] |
| μ-RWELL with DLC | ps collaboration | 10 MHz/cm$^2$ ***** | - | 35 ps * (MIP) | [35,36] |

* Measured in a test
** Expect in the final experiment (Note: $\sigma_{core} \neq \sigma_{RMS}$ in MCP single electron measurements!)
*** Status of the present running experiment
**** Expected value assuming 10-20 microns pixelization, and negligible cross-talk, charge sharing, or ASIC limitations.
***** Measured in a separate test, which had no timing measurement.

## Summary

- **MCPs**: they are the best reference-timing detectors for test beams; still attractive solution for RICH/DIRC applications. Max rate depends on choice of gain and MCP resistivity. Remaining challenges: QE aging, ALD treatment, and production reliability. The 3-MCP solution could work up to rates of 100 MHz/cm$^2$.
- **LGAD**: radiation hardness is limited by damage to the gain layer by radiation.
- **SiPM + LYSO**: limited by radiation-induced increase of SiPM dark noise.
- **3D Si**: excellent timing and high-rate capability, but presently no internal gain. My Mini-TCAD simulations indicate that hexagonal cells provide better timing resolution and better drift-time uniformity than square cells. For inclined tracks, the simulation also prefers a double threshold timing compared to a CFD method. This type of detector could work at FCC-hh collider, if it provides the gain of 5-10.
- **SiEM**: promising concept combining internal gain with potentially high radiation hardness. Timing measurements are still needed. This type of detector could work at FCC-hh collider, if it provides the gain of 5-10.
- **Gas detectors**: surprisingly good timing performance in Micromegas and μ-RWELL detectors. Various photocathodes are under study to replace CsI.

## ACKNOWLEDGEMENTS

I would like to thank A. Lehmann, S. White, A. Kiselev, C. Woody, H. Sadrozinski, E. Torassa, M. Demarteau, Z. Sýkora, K. Matsuoka, and K. Inami for contributions to this work.